\documentclass{paper}
\usepackage{ijcai20}

\usepackage{times}

\usepackage{soul}
\usepackage{url}
\usepackage[utf8]{inputenc}
\usepackage[small]{caption}
\usepackage{graphicx}
\usepackage{amsmath}
\usepackage{amsfonts}
\usepackage{booktabs}
\usepackage{amsthm}
\usepackage{multicol}
\usepackage{multirow}
\usepackage{braket}
\usepackage{array}
\usepackage{tabu}
\usepackage{bm}
\usepackage{algorithm}
\usepackage{algorithmic}
\usepackage{algorithmic}
\usepackage{epstopdf}
\usepackage{subfigure}

\title{Privacy Threats Against Federated Matrix Factorization}

\author{
Dashan Gao$^{1,2}$\footnote{Contact Author}\and
Ben Tan$^3$\and
Ce Ju$^{3}$\and
Vincent W. Zheng$^3$ \And
Qiang Yang$^{1,3}$
\\
\affiliations
$^1$Department of CSE, Hong Kong University of Science and Technology\\
$^2$Department of CSE, Southern University of Science and Technology\\
$^3$AI Lab, WeBank Co. Ltd\\
\emails
dgaoaa@connect.ust.hk,
\{btan, ceju,vincentz\}@webank.com,
qyang@cse.ust.hk
}

\begin{document}

\maketitle

\begin{abstract}
Matrix Factorization has been very successful in practical recommendation applications and e-commerce. Due to data shortage and stringent regulations, it can be hard to collect sufficient data to build performant recommender systems for a single company. 
Federated learning provides the possibility to bridge the data silos and build machine learning models without compromising privacy and security. Participants sharing common users or items collaboratively build a model over data from all the participants. There have been some works exploring the application of federated learning to recommender systems and the privacy issues in collaborative filtering systems. However, the privacy threats in federated matrix factorization are not studied.
In this paper, we categorize federated matrix factorization into three types based on the partition of feature space and analyze privacy threats against each type of federated matrix factorization model. We also discuss privacy-preserving approaches. 
As far as we are aware, this is the first study of privacy threats of the matrix factorization method in the federated learning framework.
\end{abstract}

\section{Introduction}
Recommender systems play a significant role in various applications, such as e-commerce and movie recommendation. Matrix Factorization (MF) ~\cite{koren2009matrix}, as a typical Collaborative Filter (CF) method, has positioned itself as one of the effective means of generating recommendations and is widely adopted in real-world applications. 
Traditionally, for one company, it is essential to accumulate sufficient personal rating data to build a performant MF model. However, due to the sparse nature of user-item interactions, it can be hard for a single company to collect sufficient data to build an MF model. Moreover, recently enacted stringent laws and regulations such as General Data Protection Regulation (GDPR) ~\cite{Albrecht2016HowTG} and California Consumer Privacy Act (CCPA) ~\cite{ghosh2018you} stipulate rules on data sharing among companies and organizations, making collaboration between companies by sharing personal rating data illegal and impractical. 

To tackle the challenge of protecting individual privacy and remitting the data shortage issue, federated learning (FL) ~\cite{konevcny2016federated,mcmahan2016communication} provides a promising way that enables different parties collaboratively build a machine learning model without exposing private data in each party. It addresses data silos and privacy problems together. 
In FL, data can be partitioned horizontally (example-partitioned) or vertically (feature-partitioned) into different parties. When records are not aligned between parties and the feature spaces among parties are heterogeneous, federated transfer learning can be adopted.
The use of FL in recommender systems has been studied over different data distributions. For example, ~\cite{chai2019secure} considers horizontally partitioned rating data among clients, which hold ratings of the same user-item interaction matrix. Federated multi-view MF is studied where participants hold item interaction data, item features, or user features~\cite{flanagan2020federated}. Each participant holds a part of model parameters, while some common parameters are shared among participants. 
Existing studies generally categorize horizontal and vertical federated recommender systems regarding on whether user alignment is required before FL ~\cite{yang2019flbook,chai2019secure}. For example, participants sharing different users and the same set of items implies \textit{horizontal federated recommender systems}. In our paper, we categorize different settings based on the partition of feature space, as shown in Fig. \ref{Fig.sparse_data_partition}, which is consistent with other FL systems.

Although most existing studies of federated recommender systems adopt privacy preserving techniques, including homomorphic encryption (HE)~\cite{Paillier99public-keycryptosystems}, secure multiparty computation (MPC)~\cite{DBLP:journals/iacr/MohasselZ17} and differential privacy (DP)~\cite{Dwork:2008:DPS:1791834.1791836} to protect data privacy. 
There is little exploration of how data privacy can be breached in federated MF. 
It is shown that the gradient present in the global model is the potential to breach data privacy in horizontal FL for deep learning ~\cite{Melis_2019} or logistic regression ~\cite{li2019quantification}. 
However, a comprehensive study of the privacy threats against the plaintext federated matrix factorization in different data partitions is still required. 

Inspired by this research gap, we investigate the potential privacy risks in federated matrix factorization. Specifically, we classify the federated MF into horizontal, vertical federated MF as well as federated transfer MF based on the data partition approaches. We then demonstrate how private user preference data can be breached during FL training to honest and curious participants. Finally, we discuss how cryptographic techniques can be adopted to protect privacy. 
The contribution of this paper can be summarised as follows:
\begin{itemize}
    \item We identify and formulate three types of federated MF problems based on the way of data partition. 
    \item We demonstrate the privacy threats in each type of federated MF, by showing how the private preference data can be leaked to honest-but-curious participants.
    \item We investigate privacy-preserving approaches to protect privacy in federated MF. 
\end{itemize}
In the following parts of this paper, section \ref{relatedwork} gives related works of privacy issues of MF method and federated MF. We give backgrounds of MF and security model in section \ref{background}. Then, section \ref{fedmf_privacy_threats} gives privacy attacks to each type of federated MF. In section \ref{priv_prev}, we discuss privacy-preserving approaches. Finally, conclusions are drawn in section \ref{conclusion}.

\section{Related Work}\label{relatedwork}
Privacy risks in general recommender systems are studied in ~\cite{Jeckmans2013PrivacyIR,Lam2006DoYT}, which analysis the privacy concerns that happened in different phases and caused by different entities. However, the federated learning framework is not considered, which introduces parameter exchanges between participants and thus enlarges the attack surface. Though ~\cite{chai2019secure} investigates privacy risks in horizontal federated MF and adopts HE to harness the privacy risks, it assumes honest-and-curious participants, and the proposed approach only defenses against an honest-but-curious server. We demonstrate that other curious clients can easily infer private training data. 

Many works are exploring privacy-preserving techniques for federated recommender systems. ~\cite{qi2020fedrec} adopts local differential privacy to train a neural network in horizontal federated news recommendation.  ~\cite{flanagan2020federated} studies federated multi-view MF over heterogeneous data held by different participants. ~\cite{chen2020practical} explores a two-tiered notion of privacy by introducing a set of public users.  ~\cite{canny2002collaborative} proposes federated CF based on partial Singular Value Decomposition and adopts HE for model aggregation. Fully HE ~\cite{kim2016efficient} as well as garbled circuit~\cite{nikolaenko2013privacy} is also investigated for privacy-preserving MF, where secure MF is conducted between the server and a crypt-service provider. 
~\cite{gao2019hhhfl,ju2020federated} are the first to investigate the feasibility of the FL framework to enable a distributed training of deep models from multiple heterogeneous datasets for brain-computer interface. 
Although these works propose various privacy-preserving approaches, they fail to investigate the potential privacy loss the intermediate transferred parameters can breach during FL training.

\section{Background}\label{background}
In this section, we introduce the matrix factorization method based on stochastic gradient descent, as well as the security model considered in this paper. 

\subsection{Matrix Factorization}
We consider $n$ users rate a subset of $m$ items. For $[n] := {1,\dots,n}$ the set of $n$ users, and $[m] := {1, \dots, m}$ the set of $m$ items, the user/item pairs that generate the ratings are denoted by $\mathcal{M} = \subset [n] \times [m]$. The total number of ratings is $M = |\mathcal{M}|$. Finally, for $(i, j) \in \mathcal{M}$, we denote by $r_{i,j} \in \mathbb{R}$ the rating generated by user $i$ for item $j$. Matrix factorization uses a $d \in \mathbb{N}$ dimensional vector to represent a user as $u$ and an item as $v$, referred as a profile, and models the relevance of an item to a user as the inner product of their profiles. MF computes the user profiles and item profiles $u_i, v_j \in \mathbb{R}^d$ following the \textit{regularized mean squared error} as follows:
\begin{equation}\label{mf_optimization}
    \min_{U, V} \frac{1}{M} \sum_{(i,j) \in \mathcal{M}} (r_{ij} - \langle {u_i}, v_j\rangle)^2 + \lambda_u \sum_{i\in [u]} ||u_i||_2^2 + \lambda_v \sum_{j\in [m]} ||v_j||_2^2
\end{equation}
for positive constants $\lambda_u, \lambda_v$. the inner product $\langle {u_i}, v_j \rangle$ is the predicted unobserved ratings $r_{ij}$.

Stochastic gradient descent (SGD) is widely applied to optimize the profiles $u_i$ and $v_j$ as follows:
\begin{align}
    u_i^t = u_i^{t-1} - \gamma \cdot (\nabla_{u_i} F(U^{t-1},V^{t-1}) + 2 \lambda_u u_i^{t-1}) \\
    v_j^t = v_j^{t-1} - \gamma \cdot (\nabla_{v_j} F(U^{t-1}, V^{t-1}) + 2 \lambda_v v_j^{t-1})
\end{align}
where $\gamma>0$ is the learning rate. $U$ and $V$ are the user profile matrix and item profile matrix with each row as a profile, and $\nabla_{u_i}F(U,V)$ and $\nabla_{v_j}F(U,V)$ can be computed as follows:
\begin{align}
    \nabla_{u_i}F(U,V) = - \frac{2}{M_{u_i}} \sum_{j:(i,j)\in \mathcal{M}} v_j (r_{ij} - \langle{u_i}, v_j\rangle) \label{gradient_U}\\
    \nabla_{v_j}F(U,V) = - \frac{2}{M_{v_j}} \sum_{i:(i,j)\in \mathcal{M}} u_i (r_{ij} - \langle{u_i}, v_j\rangle) \label{gradient_V}
\end{align}

\subsection{Security Model}
We assume all participants as well as the server if there is any are \textit{honest-but-curious} (a.k.a. semi-honest). An honest-but-curious participant follows the protocol honestly but tries to infer private information from the intermediate information it knows. 

\section{Federated MF and Privacy Threats} \label{fedmf_privacy_threats}

\begin{figure*}[!htb]
\centering 
\subfigure[Horizontal Federated MF]{
\label{Fig.1.1}
\includegraphics[width=0.55\linewidth]{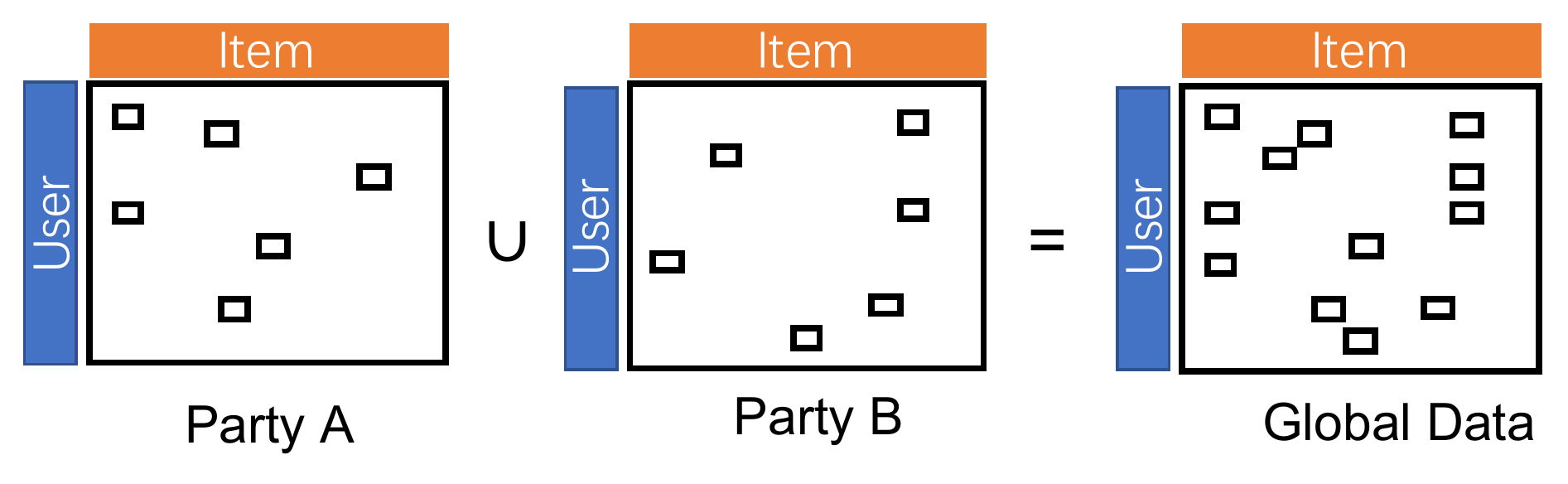}}
\subfigure[Vertical Federated MF]{
\label{Fig.1.2}
\includegraphics[width=0.34\linewidth]{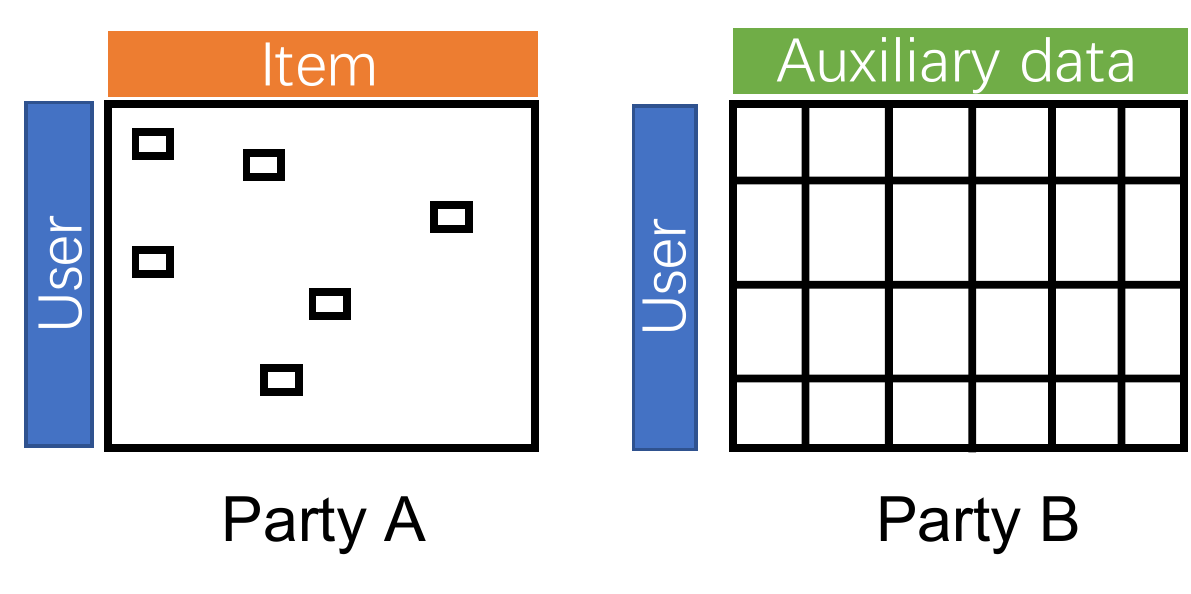}}
\subfigure[Federated Transfer MF (Common item set)]{
\label{Fig.1.3}
\includegraphics[width=0.55\linewidth]{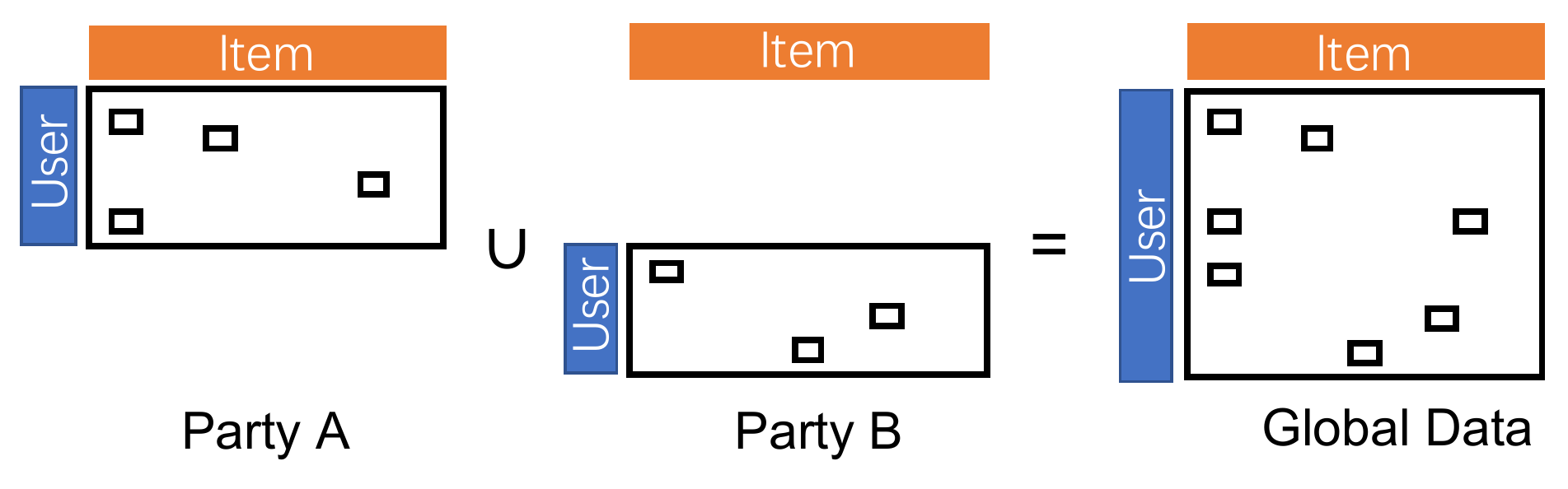}}
\caption{The way of data partition in federated MF for horizontal federated MF, vertical federated MF and federated transfer MF (assuming participants shares common item set). Orange, blue, and green rectangles denote the profile matrices of the item, user, and auxiliary data, respectively. The small rectangles in the user-item interaction matrix represent user ratings.}
\label{Fig.data_partition}
\end{figure*}

\begin{figure*}[htb]
\centering 
\subfigure[Horizontal Federated MF]{
\label{Fig.2.1}
\includegraphics[width=0.49\linewidth]{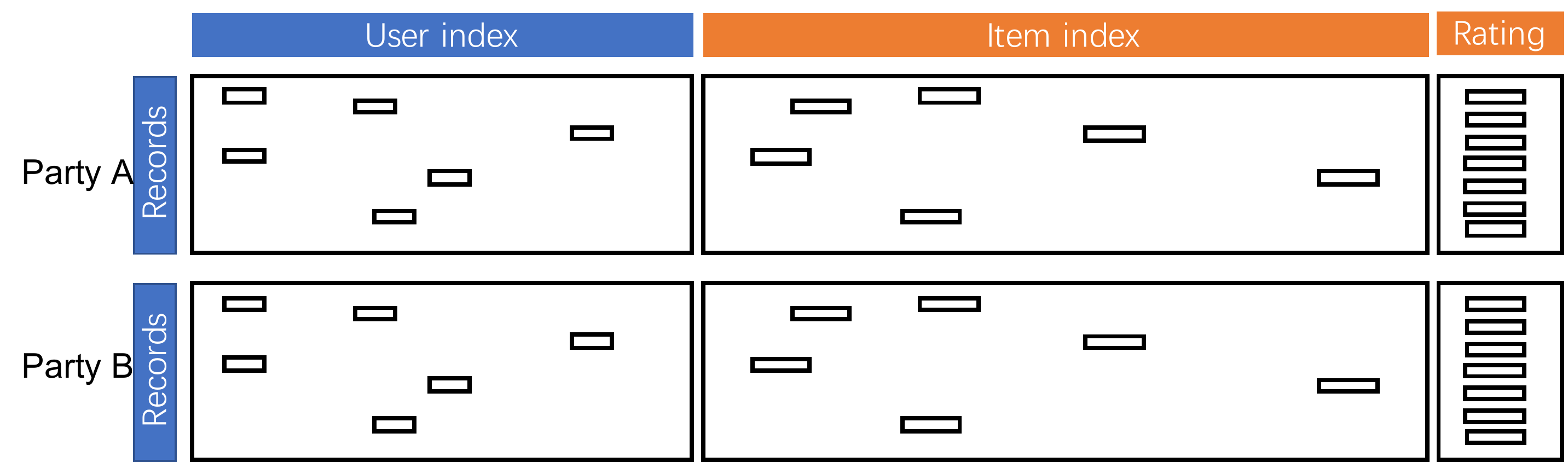}}
\subfigure[Federated Transfer MF (Common item set)]{
\label{Fig.2.2}
\includegraphics[width=0.49\linewidth]{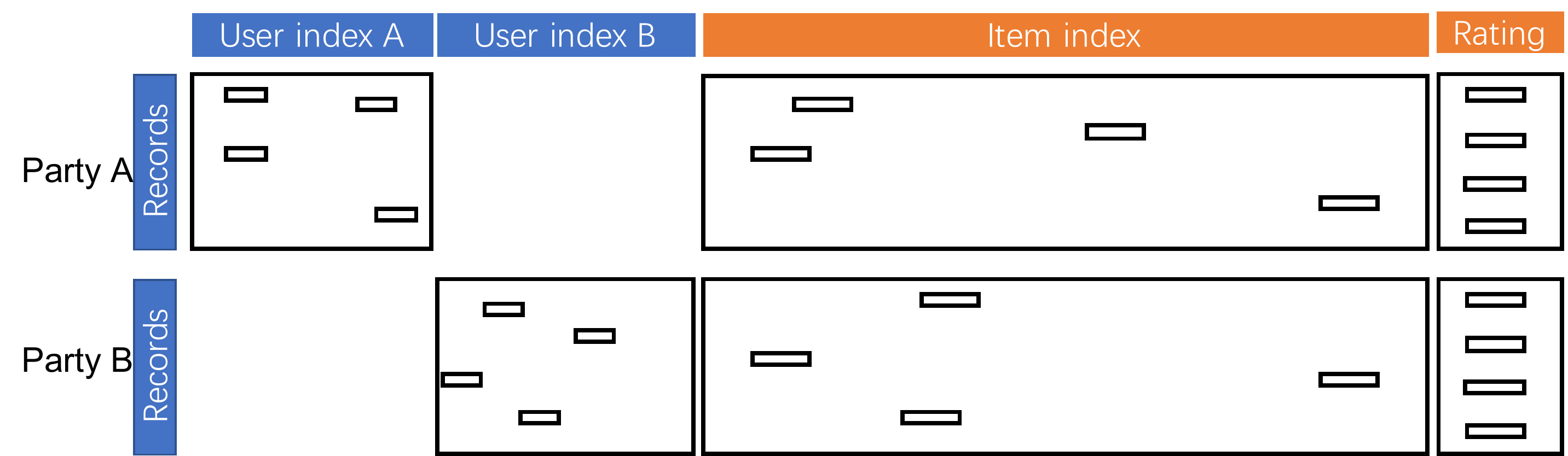}}
\subfigure[Vertical Federated MF]{
\label{Fig.2.3}
\includegraphics[width=0.48\linewidth]{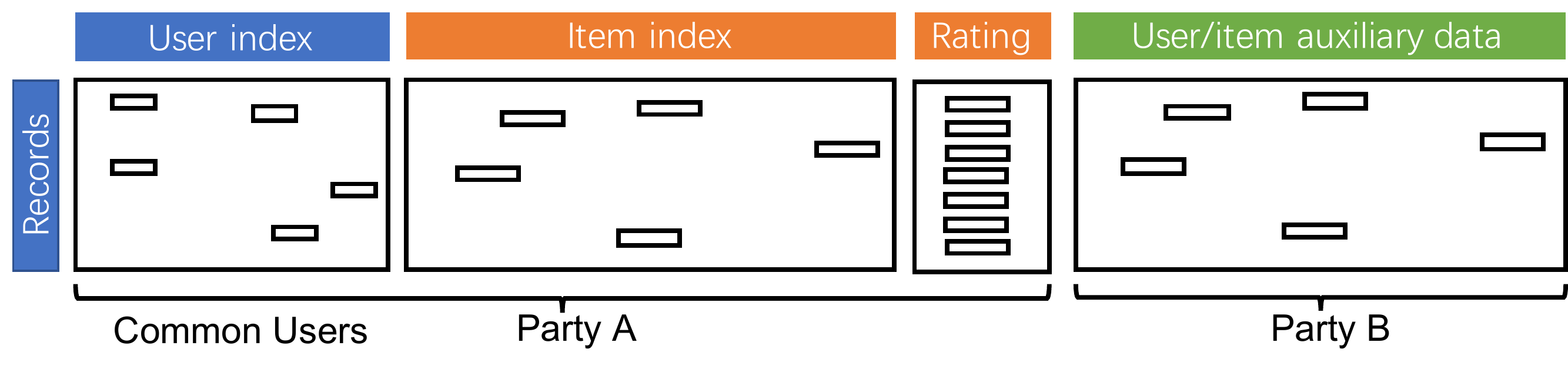}}
\caption{The sparse representation of partitioned data and feature space in federated MF for horizontal federated MF, vertical federated MF, and federated transfer MF (assuming participants share common item set). Orange, blue, and green rectangles denote the profile matrices of the item, user, and auxiliary data, respectively. The small rectangles represent non-zero values.}
\label{Fig.sparse_data_partition}
\end{figure*}

In this section, we discuss federated MF in three settings that differ in data partition. Then we investigate how privacy can be breached towards adversarial participants in each setting. For simplicity and without loss of generality, we consider FL systems consisting of two participants, $P^{A}$ and ${P^B}$. 
Fig. \ref{Fig.data_partition} compares the data partition in each setting of federated MF discussed in this paper. 
We adopt a sparse representation of partitioned data in Fig.\ref{Fig.sparse_data_partition}, to demonstrate the nature of horizontal, vertical, and transfer federated learning. In the horizontal FL setting Fig. \ref{Fig.2.1}, participants share the same feature space. In the vertical FL setting Fig.\ref{Fig.2.3}, participants hold heterogeneous feature space, and only $P^A$ holds the ratings. Whereas, in federated transfer MF Fig.\ref{Fig.2.2}, participants share partial models (e.g., item profiles) for knowledge transfer.

\subsection{Horizontal Federated MF}
In horizontal federated MF, $P^A$ and $P^B$ share the same user-item interaction matrix (i.e., the same user and item feature space), as shown in Fig \ref{Fig.1.1} and Fig. \ref{Fig.2.1}. Therefore, each participant holds the profiles of all users and items, and can locally compute gradient of the whole MF model. Only model aggregation requires communication between A and B~\cite{mcmahan2016communication}. In model aggregation, the global user profiles matrix is computed by $U_{Global} = \frac{1}{2} (U_{P^A} + U_{P^B})$, and item profiles matrix $V_{Global} = \frac{1}{2} (V_{P^A} + V_{P^B})$. $P^A$ can compute the gradient $\nabla_{U}F_B(U,V)$ of $P^B$ following
\begin{align*}
    & \nabla_{U}F_B(U^{T-1},V^{T-1}) \\
    & = \frac{1}{\gamma} (2 \cdot U_{Global}^{T} - U_{P^A}^{T} - U^{T-1}) - 2 \lambda_u U^{T-1}
\end{align*}
where $T$ is the index of round. Since the user-item interaction matrix is sparse, that is, $M = \Theta(n+m)$, which is much smaller than the number of potential ratings $n \cdot m$ ~\cite{nikolaenko2013privacy}. For one update in SGD, it is very likely to have one rating record for each item or user. Therefore, according to Equation \ref{gradient_U}, $P^A$ can easily find the $(i,j)$ pair and the corresponding $v_j$, by checking the gradient and comparing $\nabla_{u_i}F_B(U,V)$ to each of $\{u_i\}_{i=1}^{n}$ as well as comparing $\nabla_{v_j}F_B(U,V)$ to each of $\{u_j\}_{j=1}^m$. Then, $P^A$ further infer the private rating score by
\begin{equation*}
    \hat{r}_{ij} = - \frac{\nabla_{u_i}F_B(U^{T-1},V^{T-1})}{2\cdot v_j^{T-1}} + \langle {u_i^{T-1}}, v_j^{T-1} \rangle.
\end{equation*}
This way, $P^A$ may complete inference attack and extracts raw private user preference data $(i, j, \hat{r}_{ij})$ of $P^B$ from the plaintext global model in horizontal federated MF. 

\subsection{Vertical Federated MF}
In vertical federated MF as shown in Fig. \ref{Fig.1.2} and Fig. \ref{Fig.2.3}, $P^A$ holds the user-item interaction matrix, and $P^B$ holds some auxiliary data of users (or items). We adopt the model presented in ~\cite{5197422} to leverage auxiliary data provided by $P^B$ in vertical federated MF. 
For each user $i$, $P^B$ hold distinct factor vectors $y_a \in \mathbb{R}^d$ corresponds to each attribute. The user $i$ can thus be described through the set of user-associated attributes $A(u)$ as $\sum_{a\in A(u)} y_a$.
For vertical federated MF model, Equation \ref{mf_optimization} can be modified as follows:
\begin{align*}
        & \min_{U, V} \frac{1}{M} \sum_{(i,j) \in \mathcal{M}} (r_{ij} - \langle (u_i + k_i), {v_j}\rangle)^2 + \\
        & \lambda_u \sum_{i\in [u]} ||u_i||_2^2 + \lambda_v \sum_{j\in [m]} ||v_j||_2^2,
\end{align*}
where $k_i = |N(u)|^{-0.5}\sum_{l\in N(u)} x_l \sum_{a\in A(u)}$. $y_a$ is the auxiliary information of user $i$. $x$ is the implicitly preferred item set, $y$ is $i$’s attributes (e.g., demographic info).

To conduct federated vertical MF, $P^B$ locally computes and sends $k_i$ to $P^A$, and $P^A$ sends nothing to $P^B$. Therefore, $P^A$ has no privacy leakage to $P^B$, while $P^B$ leaks $k_u$ to $P^A$. In such a setting, user ID leakage during the user alignment stage causes a major privacy threat.

\subsection{Federated Transfer MF}
Without loss of generality, in federated transfer MF, we assume $P^A$ and $P^B$ holds ratings given by different users on the same set of items, and $P^A$ tries to infer private data of $P^B$. That is, $P^A$ and $P^B$ holds the item profiles matrix $V$. $P^A$ holds $U^A$, and $P^B$ holds $U^A$ for their users, respectively. To train the model, each participant locally conducts SGD to update local models. For model aggregation, only $V$ is aggregated by the participants. 
$P^A$ can learn the gradient $\nabla_V F_B(U^B,V)$ of $P^B$ as follows:
\begin{align*}
    & \nabla_{V}F_B({U^B}^{T-1},V^{T-1}) \\
    & = \frac{1}{\gamma} (2 \cdot V_{Global}^{T} - V_{P^A}^{T} - V^{T-1}) - 2 \lambda_v V^{T-1}
\end{align*}
As the interaction matrix is sparse for each round, it is reasonable to assume $\nabla_{V}F_B({U^B}^{T-1},V^{T-1}) = -2 {u^B_i}^{T-1} (r_{ij} - \langle {u^B_i}^{T-1}, {v_j}^{T-1}\rangle)$ for a user $i$. 
By collecting the gradients of several steps and assuming the $u_i^B$ does not change, which is reasonable, then the model is nearly converged, we can use some iterative methods such as Newton's method to approximate the numeric value of $u^B_i$, as shown in ~\cite{chai2019secure}. After computing $u^B_i$, the reconstructed rating score $\hat{r^B}_{ij}$ of $P^B$ can easily be computed as follows:
\begin{equation*}
    \hat{r^B}_{ij} = - \frac{\nabla_{v_j}F_B({U^B}^{T-1},V^{T-1})}{2 \cdot {u^B}_i^{T-1}} + \langle {{u^B_i}^{T-1}}, v_j^{T-1} \rangle.
\end{equation*}
Thus, $P^A$ completes the inference attack and extracts user profiles and the corresponding ratings of $P^B$ from the plaintext global model in horizontal federated MF. However, as the users are not aligned between participants, the user ID information can not be inferred by $P^A$. 

It is worth noting that, although some works denote participants sharing the same set of items and a different set of users as \textit{horizontal federated recommender systems}, and denote participants sharing the same set of users and a different set of items as \textit{vertical federated recommender systems}, based on whether the users should be aligned before FL training. Such setting demonstrates the nature of federated transfer learning, where neither the global model is shared among participants, nor the feature space is fully partitioned without intersection. Privacy attacks on both settings are the same during FL training.
Therefore, we denote both settings as federated transfer MF in this paper. 

\subsection{Comparison}\label{Comparison}
\begin{table*}[!htb]
\centering
\footnotesize
    \begin{tabular}{| c || c | c | c |}
    \hline 
    Problem setting & Parameter partition in each party & Gradient computation & Resilience against inference attack \\
    \hline 
    \hline
    Horizontal FedMF & The whole model params & Locally & Weak \\
    \hline
    Vertical FedMF & Partial params without shared params & Collaboratively & P$^A$ strong, P$^B$ weak \\
    \hline
    Federated Transfer MF & Partial params with shared params & Locally & Medium \\
    \hline
\end{tabular}
\caption{Comparison of different problem settings including horizontal and vertical federated MF as well as federated transfer MF. FedMF denotes federated MF.}\label{table_comparison}
\end{table*}

Tab. \ref{table_comparison} demonstrates the comparison of three settings based on the way to update the model, the partition of model parameters, and the resilience of the FL system against inference attack. 
In horizontal federated MF, all participants share ratings from the same set of users and items. Therefore, each participant locally holds the whole user profiles matrix and item profiles matrix for local SGD. For federated transfer MF, participants only share the same set of users (or items), each participant locally holds its user (or item) profiles sub-matrix and the global item (or user) profiles matrix for local SGD. For vertical federated MF, one party holds rating data; the other holds auxiliary data, each party holds partial parameters with shared parameters such as user profiles matrix.
For both horizontal federated MF and Federated transfer MF, clients can locally conduct SGD optimization without the need for communication. Participants only need to exchange parameters during the model aggregation process. For vertical federated MF, two participants need to collaboratively compute the estimated rating for each update, which dramatically increases the communication cost. The resilience of each setting against the inference attack is also shown. Horizontal federated MF breaches most private information, including user ID and user preference data. For vertical federated MF, recommender $P^A$ leaks no information to data provider $P^B$, and data provider sends the intermediate data to the recommender. The user ID is breached for both participants. For federated transfer MF, only private rating data and user profiles are leaked, and no user ID is breached.

\section{Privacy Preservation in Federated MF}\label{priv_prev}
According to the privacy threats investigated in section \ref{fedmf_privacy_threats}, we give some advises for privacy preservation in federated MF.
For horizontal federated MF, the global user and item profile matrices computed by aggregation should be protected against each participant. For vertical federated MF, the auxiliary data provider should keeping its computed feature sent to the recommender secret. For federated transfer MF, the shared user or item profile matrix should be kept secret to any honest-but-curious participant throughout the FL training, as the rating score and private profile can be potentially implied.

To keep intermediate parameters private, there are mainly three types of approaches \textit{cryptography-based}, \textit{obfuscation-based} and \textit{hardware-based} approaches. Cryptography-based approaches generally use HE and MPC to keep intermediate transactions private. Obfuscation-based approaches such as DP obfuscate private data by randomization, generalization or repression. Hardware-based approaches rely on trusted execution environment (TEE) to conduct FL learning in a trusted enclave.
By using cryptography-based approaches, fully HE can be introduced to prevent decryption during training~\cite{kim2016efficient}. Secret sharing schemes can also be introduced following a two-server architecture~\cite{cryptoeprint:2011:535}. 
Since the user-item interaction matrix is sparse, applying DP may introduce too much noise and make the model unavailable. TEE can also be applied by encrypting private data and conducting private training inside TEE~\cite{CHEN202069}.

\section{Conclusion}\label{conclusion}
We identify and formulate three types of federated MF problems based on the partition of feature space. Then, we demonstrate the privacy threats against each type of federated MF. We show how the private user preference data, private user/item profiles matrix, and user ID can be potentially leaked to honest-but-curious participants. Finally, We discuss privacy-preserving approaches to protect privacy in federated MF. 
For future work, we will experimentally study the power of the proposed privacy attacks by measuring the portion and accuracy of the inferred private data. Privacy threats against alternating least squares-based MF and other recommender systems also require further comprehensive study.

\bibliographystyle{named}
\bibliography{ijcai20}

\end{document}